\documentclass[sigconf]{acmart}

\usepackage{bm} 
\usepackage{amsthm}
\usepackage{caption}
\usepackage{subcaption}
\usepackage{algorithm}
\usepackage[noend]{algpseudocode}
\usepackage{thmtools}
\usepackage{enumitem}
\usepackage{tabularx}
\usepackage{xcolor} 
\usepackage{lipsum}
\usepackage{booktabs}
\usepackage{multirow}

\newtheorem{defn}{Definition}
\newtheorem{theorem}{Theorem}
\newtheorem{proposition}{Proposition}
\newtheorem{lemma}{Lemma}
\newcolumntype{L}[1]{>{\raggedright\arraybackslash}p{#1}}
\newcolumntype{C}[1]{>{\centering\arraybackslash}p{#1}}
\newcolumntype{R}[1]{>{\raggedleft\arraybackslash}p{#1}}

\AtBeginDocument{%
  \providecommand\BibTeX{{%
    \normalfont B\kern-0.5em{\scshape i\kern-0.25em b}\kern-0.8em\TeX}}}

\setcopyright{acmcopyright}
\copyrightyear{2018}
\renewcommand\footnotetextcopyrightpermission[1]{} 

\acmConference[WWW '21]{Woodstock '18: ACM Symposium on Neural
  Gaze Detection}{April 19--23, 2021}{Ljubljana, Slovenia}
\acmBooktitle{Woodstock '18: ACM Symposium on Neural Gaze Detection,
  June 03--05, 2018, Woodstock, NY}
\acmPrice{15.00}
\acmISBN{978-1-4503-XXXX-X/18/06}

\begin{document}

\title{GAN-based Recommendation with Positive-Unlabeled Sampling}








\author{Yao Zhou$^\dagger$,~~~Jianpeng Xu$^\ddagger$, ~~~Jun Wu$^\dagger$,~~~Zeinab Taghavi Nasrabadi$^\ddagger$,~~~Evren Korpeoglu$^\ddagger$, \\~~~Kannan Achan$^\ddagger$,~~~Jingrui He$^\dagger$}
\affiliation{
    \institution{$^\dagger$University of Illinois at Urbana Champaign, \{yaozhou3,junwu3, jingrui\}@illinois.edu; \\ $^\S$Walmart Labs, \{Jianpeng.Xu,ZTaghavi,EKorpeoglu,KAchan\}@walmartlabs.com}
}


\newcommand{\he}[1]{{\textsf{\textcolor{red}{[From He: #1]}}}}
\newcommand{\yao}[1]{{\textsf{\textcolor{blue}{[From Yao: #1]}}}}

\begin{abstract}
Recommender systems are popular tools for information retrieval tasks on a large variety of web applications and personalized products. In this work, we propose a Generative Adversarial Network based recommendation framework using a positive-unlabeled sampling strategy. Specifically, we utilize the generator to learn the continuous distribution of user-item tuples and design the discriminator to be a binary classifier that outputs the relevance score between each user and each item. Meanwhile, positive-unlabeled sampling is applied in the learning procedure of the discriminator. Theoretical bounds regarding positive-unlabeled sampling and optimalities of convergence for the discriminators and the generators are provided. We show the effectiveness and efficiency of our framework on three publicly accessible data sets with eight ranking-based evaluation metrics in comparison with thirteen popular baselines. 
\end{abstract}





\maketitle

\section{Introduction} \label{sec:intro}
\sloppy Recommender systems have been prevalent in recent decades across multiple domains in e-Commerce~\cite{Xu2020}, content streaming (YouTube)~\cite{Covington2016}, and business service industries (Yelp)~\cite{Tay2018}, due to their success in filtering or retrieving relevant information from user profiles and behaviors. Traditional collaborative filtering methods~\cite{SlopeOne,Co-Clustering} and matrix factorization methods~\cite{SVD,NMF,PMF} are the most popular and effective set of methods of recommender systems for many years. Recently, various embedding based methods such as deep factorization machine~\cite{deepfm} and neural collaborative filtering~\cite{NCF_www} have been proposed and achieved impressive performance. This leads to a wide and in-depth study of the deep learning based recommender systems~\cite{Zhang2019}. 
Most of the existing methods take the following two assumptions for granted, especially for \textit{implicit} recommender systems: (1) The unobserved interactions between users and items (i.e., unlabeled user-item tuples) are often labeled as negative samples; (2) The observed users, items, and their interactions are representing the true relevance distribution. However, these two assumptions are usually not valid for real-world recommender systems. 

In the first assumption, it is assumed that an item $i$ is more relevant to a user $u$ than item $j$ if $i$ has interactions with $u$ while $j$ does not. The assumption is not necessarily true in that, the missing of interactions between item $j$ and user $u$ could be because of the lack of the exposure between item $j$ and user $u$, rather than the uninterestingness of $u$ on $j$. In other words, the unlabeled user-item tuple can be either a positive or negative sample. Hence, simply using the unlabeled tuples as negative samples in the training process can inevitably degrade the model performance. 
In this paper, instead of taking the unlabeled tuples as negative samples, we formulate the recommender system into a Positive-Unlabeled (PU) learning~\cite{Bekker2018} framework, which is a machine learning approach where the learner observes only positive data and unlabeled data. PU learning has been applied in a variety of applications, such as ecology~\cite{Ward2009}, healthcare~\cite{Yang2012}, and remote sensing~\cite{Li2011}.
Existing works of PU learning mainly focus on designing the PU learning adapted objectives~\cite{PUtheory}. It has been theoretically analyzed in~\cite{nnPU} that for unbiased PU learning, the empirical risks on training data can be negative if the training model is very flexible, which will result in serious overfitting. Hence, even though flexible models such as deep neural networks have been widely explored in recommender systems, limited work has been done under the PU learning setting. 

Secondly, in traditional recommender systems, the training samples are usually composed of the positive (labeled) samples and a sampled set of negative samples from the unlabeled data. This negative sampling process can be problematic in that, as we mentioned in PU learning, the samples from the unlabeled data may not necessarily be the real negative ones, and this will distort the learned data distribution in the modeling process. Generative models such as generative adversarial networks (GAN)~\cite{Goodfellow2014} tried to alleviate the issue of negative sampling by learning the underlying data distribution from an implicit generative model instead of imposing any assumption on the existing data. In the framework of GAN, a discriminator is introduced to distinguish the generated samples of the generator from the real samples, while the generator is optimized in such a way that its generated samples are hardly separable by the discriminator. Specifically, IRGAN~\cite{IRGAN} was proposed to apply GAN on learning-to-rank applications, including recommender systems. IRGAN employed policy gradient based reinforcement learning to perform discrete sampling of documents (items) for each query (user), in order to select relevant items from a given pool. However, we argue that this discrete sampling strategy may limit the expressiveness of the generator due to the sparsity of data in recommendation, and the model will not learn the underlying true distribution of the users and items. Besides, IRGAN only performed sampling on items, and used all users in the loss function, which makes IRGAN lack the capability to learn the distribution of the users. 


In order to address the aforementioned limitations of existing works, we propose a novel approach called \textbf{P}ositive-\textbf{U}nlabeled \textbf{RE}commendation with generative adversarial network (PURE). First of all, based on the analysis of \cite{nnPU}, PURE adopts the \textit{positive unlabeled risk minimizer} to train an unbiased positive-unlabeled discriminator. In particular, we theoretically prove that the estimation error bound of PU Learning is tighter than that of positive-negative (PN) learning when the number of unlabeled samples is lower bounded, which can be easily satisfied due to the extreme sparsity of the real-world data. In addition, in order to learn the true distribution of users and items, continuous sampling on both users and items in the embedding space is employed in the generator. Specifically, a fake item (embedding) for a user is generated with a random noise input. A fake user (embedding) can also be generated in a similar way. Furthermore, we theoretically prove that the optimal generator is able to generate high-quality embeddings from a learned user-item distribution that is very similar to the true user-item relevance distribution.

The main contributions of this paper are summarized below: 
\begin{itemize}
    \item We propose a novel approach for recommender systems called PURE under the GAN framework, which trains an unbiased positive-unlabeled discriminator using PU learning. 
    \item The generator of PURE performs continuous sampling on both users and items in the embedding space in order to learn the true relevance distribution of the users and items. 
    \item We theoretically prove that an unlabeled sampling bound of PURE exists and can be satisfied easily in real-world recommender systems. The optimalities upon convergence are also provided for both the discriminator and the generator.
    \item We show the effectiveness of PURE\footnote{The code is available at this anonymous link: \url{https://drive.google.com/drive/folders/1Zf_NrnBmUfYb78z8Qo7zJ_oAYzihRVqS?usp=sharing}} on three public data sets using eight ranking based evaluation metrics compared with thirteen popular baselines. 
\end{itemize}

The rest of the paper is organized as follows. Section~\ref{sec:probDef} is the preliminary. Section~\ref{sec:pugan} describes the proposed framework PURE, and Section~\ref{sec:analysis} presents the analyses of PURE from various perspectives. The experimental results are illustrated in Section~\ref{sec:experiment}. In Section \ref{sec:relatedwork}, we briefly introduce the related work on recommender systems and PU learning. In the end, we conclude the paper in Section \ref{sec:concludsion}.

\section{Preliminary} \label{sec:probDef} 
In this section, we first present the notation as well as the problem definition for recommendation. Then, the preliminary work of generalized matrix factorization (GMF) and generative adversarial network (GAN) are briefly reviewed.

\subsection{Problem Definition}
We let $\mathcal{U}$ and $\mathcal{I}$ denote the sets of users and items. Given a user $u$, a list of relevant items can be rated (explicitly) or viewed (implicitly) by $u$. From the perspective of matrix representation, we define the user-item interaction matrix as $\mathcal{R} \in \{1,0\}^{M \times N}$, where $M$ and $N$ denote the number of users and items, respectively. The entry $\mathcal{R}_{ui} = 1$ if there is an observed interaction (explicitly or implicitly) between user $u$ and item $i$. We further assume $\Omega$ to be the index set of these observed entries, namely, $(u,i) \in \Omega$ if $\mathcal{R}_{ui}=1$. It should be noticed that $\mathcal{R}_{ui}=0$ does not necessarily mean that user $u$ dislikes item $i$. The unobserved entries could be missing data with either positive labels (i.e., user and item are truly relevant) or negative labels (i.e., user and item are non-relevant). In real applications, each user can only rate and view a very limited number of items. Therefore, without loss of generality, we assume the truly relevant user-item tuples are very sparse in nature. 

Then, the recommendation problem is usually formulated as follows:
\begin{defn}
[\textbf{Recommendation Problem}] \leavevmode \\
\textbf{Given:} A set of users $\mathcal{U}=\{u_1, u_2, ..., u_M\}$, a set of items $\mathcal{I}=\{i_1, i_2, ..., i_N\}$, the observed user-item interaction matrix $\mathcal{R}$. \\
\textbf{Output:} The estimated interaction scores of the unobserved items for each user $u$ in $\mathcal{U}$.
\end{defn}

\subsection{Generalized Matrix Factorization}
Matrix Factorization (MF) is one of the most successful recommendation approaches that realize the latent factor models by decomposing the user-item matrix $\mathcal{R}$ into the product of two lower dimensional matrices. The MF model usually maps both users and items to a joint latent factor space with the dimensionality of $d$. Accordingly, each user $u$ is associated with a latent vector $\bm{e}_u \in \mathbb{R}^d$, and each item $i$ is associated with a latent vector $\bm{e}_i \in \mathbb{R}^d$. To learn these latent factor vectors, the objective is usually designed to minimize the squared error on the observed user-item tuples: 
\begin{equation}
    \min_{\{\bm{e}_u, \bm{e}_i\}} \sum_{(u,i) \in \Omega} (\mathcal{R}_{ui} - \bm{e}_u^{\top}\bm{e}_i)^2
\end{equation}
Despite its success in various applications, MF assumes user and item latent features are equally important on each dimension, and combines them with equal weights. However, \cite{NCF_www} has pointed out that MF can incur a large ranking error due to its naive assumption. Therefore, they propose to use a GMF model to increase the expressiveness of MF:
\begin{equation} \label{eq:GMF}
    \min_{\{\bm{e}_u, \bm{e}_i\}} \sum_{(u,i) \in \Omega \cup \Omega^-} \Big(\mathcal{R}_{ui} - \{\bm{e}_u \odot \bm{e}_i\}^\top \bm{r}_D \Big)^2
\end{equation}
where $\odot$ is the element-wise product and $\Omega^-$ denotes the set of negative samples, which are sampled from the unobserved user-item interactions. $\bm{r}_D$ is a learnable vector which builds the relation mapping between user latent vector $\bm{e}_u$ and item latent vector $\bm{e}_i$.

\subsection{Generative Adversarial Network}
GAN was initially introduced in \cite{Goodfellow2014} and it consists of two models, i.e., discriminator $D$ and generator $G$, that play a minimax game. The discriminator $D$ aims to distinguish the real-world data and the fake data from the generator $G$. Meanwhile, the generator $G$ aims to generate fake data to confuse the discriminator $D$ as much as possible. The objective of GAN is usually formatted as:
\begin{equation} \label{eq:ori_GAN}
    \min_{G} \max_{D} V(D,G)=\mathbb{E}_{p_{data}(x)}\Big[ \mathrm{log}D(x) \Big] + \mathbb{E}_{p_{\textit{g}}(x)}\Big[ \mathrm{log}(1-D(x)) \Big]
\end{equation}
where $p_{data}(x)$ and $p_g(x)$ represent the distributions of real-world data and generator $G$'s fake output data. The objective of GAN is equivalent to minimizing the Jensen-Shannon Divergence between $p_{data}(x)$ and $p_g(x)$. Therefore, upon convergence, we expect $G$ to generate high-quality fake data that are visually similar to the real data. The problem in Eq.~(\ref{eq:ori_GAN}) is the conceptual formulation of GAN that favors the theoretical analysis, however, in implementation, we still need to include the objective function for loss calculation and gradient back-propagation. Then, the objective becomes:
\begin{equation}
    \min_{G} \max_{D} V(D,G)=\mathbb{E}_{p_{data}(x)}\Big[ f_{D}(D(x)) \Big] + \mathbb{E}_{p_{\textit{g}}(x)}\Big[ f_{G}(D(x)) \Big]
\end{equation}
where $f_D$ and $f_G$ are the loss functions for discriminator $D$ and generator $G$, respectively.

\section{Proposed Approach} \label{sec:pugan}
\begin{figure}[!t]
\centering
\includegraphics[width=8.5cm]{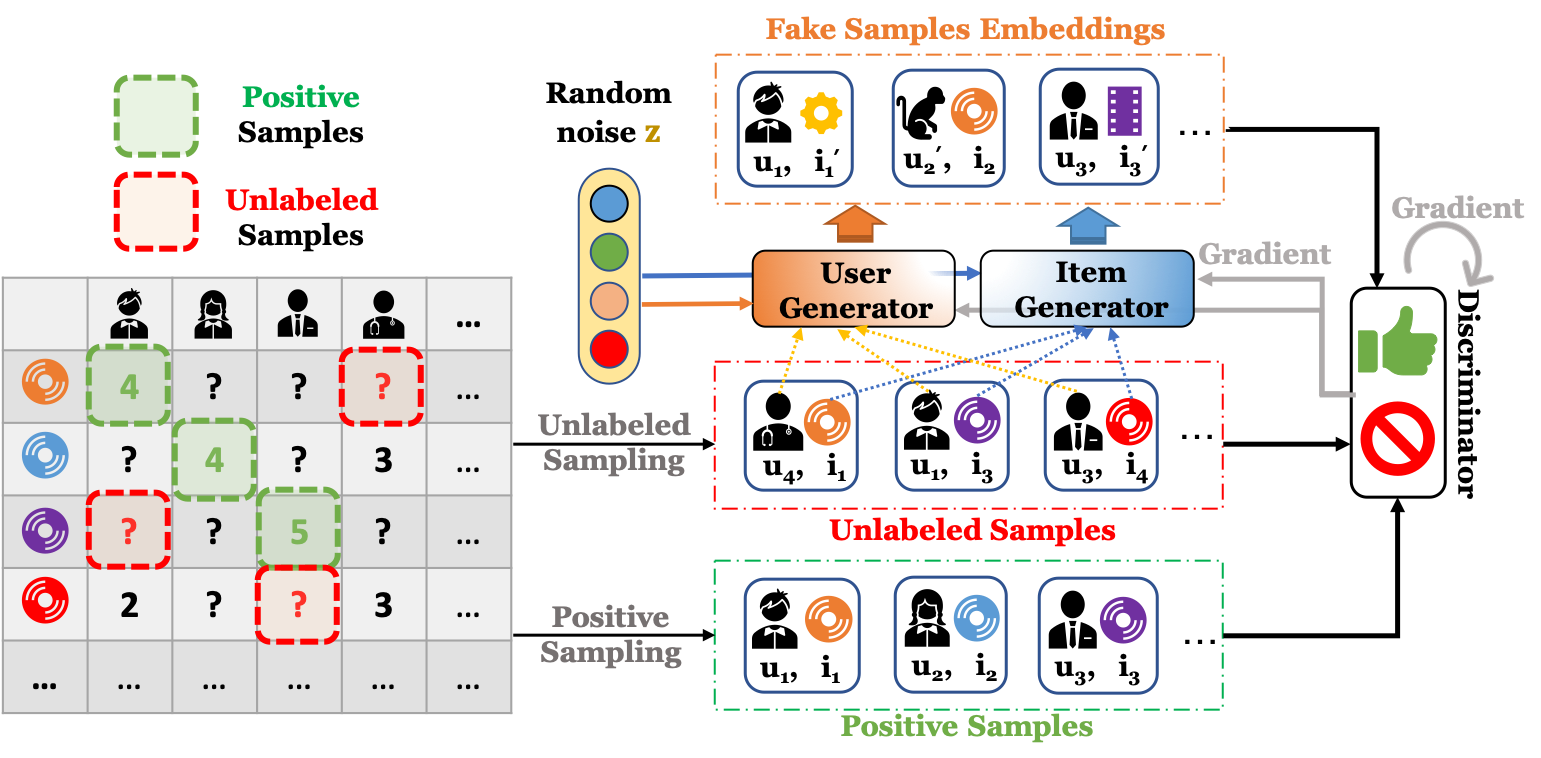}
\vspace{-6mm}
\caption{Overview of the proposed PURE framework}
\label{PUGAN-overview}
\vspace{-4mm}
\end{figure}

This section presents our proposed framework PURE. We first describe positive-unlabeled (PU) learning in the recommendation setting, emphasizing on how to learn a supervised discrimination model with PU risk estimators. Then, following the GAN framework, the discriminator in PURE has the ability to take various types of training samples into consideration, while the generator could generate the fake user and fake item embeddings that cover the corners of the continuous latent space, which increases model expressiveness. The overview of the PURE is shown in Figure~\ref{PUGAN-overview}.

\subsection{PU Classifications in Recommendation}
In recommendation, we usually learn to map each user-item tuple $(u,i)$ to a scalar value that can represent the relevance of $i$ to $u$. In our framework, we design the discriminator $D(u, i)$ to be able to maps $(u,i)$ to the value of $Y \in \{0, 1\}$. The goal of the discriminative model is to distinguish between the truly relevant items and non-relevant items for the given user. Intuitively, the discriminator $D(u,i)$ is simply a binary classifier that outputs a probability relevance score. This output score should be $1$ when the item $i$ is truly relevant to the user $u$, and should be $0$ when $u$ and $i$ are non-relevant. Formally, we quantify the output score of the discriminator as:
\begin{equation}
    D(u,i) = \frac{1}{1+\mathrm{exp}\Big(-\phi(u,i)\Big)}
\end{equation}
where we let $\phi(u,i):\mathbb{N} \times \mathbb{N} \rightarrow \mathbb{R}$ be the decision function of the discriminator $D(u, i)$ and $\mathbb{N}$ is the set of natural numbers for user and item indices. The specific instantiation of decision function $\phi(u, i)$ can be versatile (e.g., matrix factorization~\cite{SVD}, factorization machine~\cite{FM}, neural networks~\cite{NCF_www}, etc.).

We let $p_{data}(u,i)$ be the underlying joint distribution of users and items, and $\pi_p = p(Y=1)$ be the positive class prior. Then, this joint distribution can be rewritten as follows based on the law of the total probability:
\begin{equation} \label{eq:data_distribution}
    p_{data}(u,i) = \pi_p p_{p}(u,i) + (1-\pi_p) p_{n}(u,i)
\end{equation}
Here, the positive user-item tuples are assumed to be drawn from the positive marginal distribution $p_p(u,i)=p_{data}(u,i|Y=1)$, and the negative tuples are drawn from the negative marginal distribution $p_n(x)=p_{data}(u,i|Y=0)$. 

To train the recommendation model, we let $L(\hat{y},y)$ be the loss function, where $y$ is the ground truth and $\hat{y}$ is the prediction. Then, the expected learning risk of the discriminator is $R(D)=\mathbb{E}_{p_{data}(u,i)}\Big[L\Big(D(u,i),Y\Big)\Big]$. Thereby, a positive-negative (PN) risk minimizer for $D$ can be learned as:
\begin{equation} \label{PNrisk}
    \min_D R(D) = \pi_p R^+_p(D) + (1-\pi_p) R^-_{n}(D)
\end{equation}
where $R^+_p(D) = \mathbb{E}_{p_p(u,i)} \Big[ L\Big(D(u,i), 1\Big) \Big]$ is the risk of the relevant samples w.r.t. the positive labels ($Y=1$) and $R^-_n(D) = \mathbb{E}_{p_n(u,i)} \Big[L\Big(D(u,i), 0\Big)\Big]$ is the risk of the non-relevant samples w.r.t. the negative labels ($Y=0$). In practice, $R^+_p(D)$ can be approximated empirically using the observed relevant user-item tuples, but $R^-_n(D)$ is usually unknown. To estimate the learning risk, many existing work simply assume the set of the unobserved user-item tuples from the unlabeled distribution $p_u(u,i)$ are non-relevant, and perform negative sampling by assigning these tuples with negative labels.

Nevertheless, this assumption can hardly be satisfied in real scenarios since such ``negatively'' sampled data will inevitably include a certain number of positive samples. Naively assigning them with negative labels, the training process of the recommender system is usually unstable and often has poor convergence~\cite{BPR}. To this end, PU learning~\cite{nnPU,PUtheory} can be used to tackle this problem with theoretical guarantees by treating the unobserved user-item tuples directly as unlabeled samples. Following~\cite{nnPU}, we also express the unlabeled marginal distribution as $(1-\pi_p) p_n(u,i) = p_{u}(u,i) - \pi_p p_p(u,i)$. Then, $R^-_n(D)$ has the following equality:
\begin{equation}
\begin{aligned}
    (1-\pi_p)R_n^-(D) = \; R_{u}^-(D) - \pi_p R_p^-(D)
\end{aligned}
\end{equation}
where $R_{u}^-(D) = \mathbb{E}_{p_{u}(u,i)}\Big[L\Big(D(u,i), 0\Big)\Big]$ is the risk of unlabeled samples w.r.t. the negative labels, and $R_p^-(D) = \mathbb{E}_{p_p(u,i)}\Big[L\Big(D(u,i), 0\Big)\Big]$ is the risk of positive samples w.r.t. the negative labels. Thus, the final risk minimization problem can be rewritten as:
\begin{equation} \label{PUrisk}
    \min_D R(D) = \pi_p R^+_p(D) - \pi_p R^-_p(D) + R^-_u(D)
\end{equation}
By minimizing the objective of Eq.~(\ref{PUrisk}), the discriminator $D$ can distinguish the relevance of user-item tuples by minimizing the learning risks of $p_p(u,i)$ and $p_u(u,i)$. Note that due to the negative property of the second term in Eq.~(\ref{PUrisk}), many existing work~\cite{nnPU,pugan_cvpr} may replace it with $\mathrm{max}\{0, - \pi_p R^-_p(D) + R^-_u(D)\}$ to guarantee a non-negative risk. However, in recommendation, the positive class prior $\pi_p$ is always very small which alleviates this issue, and we did not observe such a negative risk phenomenon in our experiments without adding the max operator.

\subsection{Discriminative Model}
With the well-defined risk minimization objective, now we demonstrate how to empirically train the discriminator using the following sets of training samples:

\noindent\textbf{Positive samples from given observations}. User $u$ and item $i$ are observed in the given data set and are truly relevant ($\mathcal{R}_{ui}=1$). For these samples, the discriminator aims to maximize the following objective:
\begin{equation} \label{D_loss1}
    V(D)_1 = \sum_{(u,i) \in \Omega}^{n_p} \pi_p \mathrm{log} D(u,i) - \pi_p \mathrm{log} \Big(1-D(u,i)\Big) 
\end{equation}
where $n_p=|\mathcal{R}|$ is the number of observed positive tuples. To comply with the PU learning objective in Eq.~(\ref{PUrisk}), the second term is the empirical risk of positive samples w.r.t. negative labels. Intuitively, we want to maximize (minimize) the $D$'s predictions on samples with positive (negative) labels. 

\noindent\textbf{Unlabeled samples from unobserved interactions and the generator}. Given a user $u$, the discriminative model is designed to assign lower scores to the items that have not be rated or viewed. We decompose this part of the objective from both the unobserved samples and generated user-item samples:
\begin{equation} \label{D_loss2}
        V(D)_2= \sum\limits_{(u,i) \in \Omega^-}^{n_u} \mathrm{log}\Big(1-D(u,i)\Big)
        + \Big[\mathrm{log}\Big(1-D(u,i')\Big) + \mathrm{log}\Big(1-D(u',i)\Big)\Big]
\end{equation}
where the fake user $u'\sim G(z_u)$ and fake item $i'\sim G(z_i)$ are generated from the user and item generators respectively, and $n_u$ is the number of unlabeled tuples from unlabeled sampling. The ratio between the unlabeled samples generated by the generator and sampled from unobserved tuples could be a hyper-parameter to tune. Here, we set their ratio to be $1$ in the experiments, namely, these two sources of unlabeled samples are equally important. However, further tuning of this ratio may lead to better performance, and we leave it for future exploration.

\subsection{Generative Model} 
The generative model aims to generate fake samples to fool the discriminator as much as possible. Therefore, given a real sample $(u, i)$, the generator $G_i(z_i)$ is designed to generate a fake item $i'$ that is highly likely to be relevant to $u$. This fake item can be virtual, and do not even exist in $\mathcal{I}$. Similarly, the generator $G_u(z_u)$ will generate a fake user $u'$ that is likely to be relevant to $i$. In particular, we design the noise input for user and item generators to be a random Gaussian noise:
\begin{equation} \label{eq:noise_input}
    z_i, z_u \sim \mathcal{N}(\bm{0}, \delta \bm{I}) 
\end{equation}
where the mean of noise input would be a zero vector $\bm{0}$ of the same size as embedding dimension $d$, and $\bm{I} \in \mathbb{R}^{d \times d}$ is the identity matrix whose magnitude is controlled by $\delta$ which represents the underlying deviations of the generator's noise input. Next, we apply the multi-layer perceptron (MLP) to generate the fake item $i'$ and user $u'$ as follows:
\begin{equation} \label{generator}
\begin{split}
    i' \sim G_i(z_i) &= {\tt ReLU}\Big( W_i^2 \cdot {\tt ReLU}\Big( W_i^1 \cdot z_i + b_i^1)\Big) + b_i^2 \Big) \\
    u' \sim G_u(z_u) &= {\tt ReLU}\Big( W_u^2 \cdot {\tt ReLU}\Big( W_u^1 \cdot z_u + b_u^1)\Big) + b_u^2 \Big)
\end{split}
\end{equation} 
where $W_i^1, W_i^2$ and $b_i^1, b_i^2$ are the learnable weights and biases for the $1$-st layer and the $2$-nd layer of MLP in the item generator $G_i(z_i)$, and we have similar definitions for the user generator $G_u(z_u)$. In the experiments, we observe that a two-layer MLP would be very effective and computationally efficient. Then, putting everything together, we have the overall objective of PURE as follows:
\begin{equation} \label{eq:final_objective}
\resizebox{1.1\hsize}{!}{$
\begin{aligned}
    &\min_{G} \max_{D} V(D,G) = \; \sum_{(u,i) \in \Omega}^{n_p} \pi_p \mathrm{log} D(u,i) - \pi_p \mathrm{log} \Big(1-D(u,i)\Big) \\
    +& \sum\limits_{(u,i) \in \Omega^-}^{n_u} \mathrm{log}\Big(1-D(u,i)\Big) + \Big[\mathrm{log}\Big(1-D\big(u,G_i(z_i)\big)\Big) + \mathrm{log}\Big(1-D\big(G_u(z_u),i\big)\Big)\Big]
\end{aligned}
$}
\end{equation}

The above objective can be optimized by performing a gradient-based optimization method. We find that Adam~\cite{Adam} would be empirically more stable and converge faster than other optimizers.


\section{Model Analysis}\label{sec:analysis}
In this section, we analyze the performance of the proposed PURE framework from multiple aspects.

\subsection{Instantiation of the Discriminator}
For discriminator's decision function $\phi(u,i)$, we can define it in various ways. In our experiment, we adopt the design of GMF (see Eq.~(\ref{eq:GMF})) by assuming that user and item embeddings have the same dimensionality:
\begin{equation}
    D(u,i) = \frac{1}{1+\mathrm{exp}(-\{\bm{e}_u \odot \bm{e}_i\}^{\top} \bm{r}_D)}
\end{equation}
In practice, we set the user embedding and item embedding to have the same dimension. Nevertheless, it is rather straightforward to extend it to a more general setting that users and items have different embedding dimensions. Then, a more generalized form for quantifying the output score of discriminator $D$ is: 
\begin{equation} \label{eq:generalized_D}
    D(u,i) = \frac{1}{1+\mathrm{exp}(-\bm{e}_u^{\top} M_D \bm{e}_i)}
\end{equation}
where $\bm{e}_u \in \mathbb{R}^{d_u}$ and $\bm{e}_i \in \mathbb{R}^{d_i}$ are the latent embeddings of user $u$ with size $d_u$ and item $i$ with size $d_i$, respectively. $M_D \in \mathbb{R}^{d_u \times d_i}$ is the learnable relation mapping matrix for user and item embeddings. Note that MF-based and GMF-based discriminators are both special cases of Eq.~(\ref{eq:generalized_D}) by setting $d=d_u = d_i$ and $M_D$ as an identity matrix or a diagonal matrix.

\subsection{Sampling Strategy}
In PN learning, it is a common practice to treat the observed user-item tuples as positive, and treat the rest as negative. However, due to the sparsity of the positive tuples, we frequently sample the negative tuples from a large number of unlabeled entries. One popular sampling strategy is uniform negative sampling (UNS), where the number of sampled ``negative'' tuples $n_n$ is proportional to the number of positive tuples $n_p$. Nevertheless, UNS may lead to poor and unstable convergence~\cite{BPR} during training due to its ill-conditioned assumption. To stabilize and improve the model performance, other techniques have been developed to alleviate the convergence issue, such as dynamic negative sampling (DNS) or dynamic random negative sampling (DRNS)~\cite{IRGAN}. Their intuitions are similar to the concept of one-class SVM~\cite{OneClassSVM}, which wraps a classification boundary around the positive samples and treats the rest as negative. 
Both DNS and DRNS have been shown to be faster in terms of model convergence~\cite{BurgesSRLDHH05, BPR, IRGAN} in the PN learning setting. However, both of them need to call the learned model repeatedly which is extremely computational expensive especially for large-scale data sets. In PURE, we adopt the efficient UNS sampling strategy since unlabeled data have been explicitly modeled in our PU learning objective. 

\subsection{Sampling Bound}
Another key question is how to determine the number of unlabeled samples $n_u$. In PN learning, the selection of $n_n$ is usually empirical, where  $n_n = C n_p$ and $C$ is the negative sampling ratio. 
However, in PU learning, with the utilization of estimation error bound~\cite{PUtheory}, $n_u$ can be determined by $\pi_p$ and $n_p$ using the following theorem.
\begin{theorem} \label{theorem:sampling_bound}
[\textbf{Unlabeled Sampling}] The estimation error bound of PU learning is tighter than that of PN learning if and only if:
\begin{equation} \label{eq:PUsampling}
    n_u \geq \frac{\sqrt{C} \, n_p}{\Big(1-\big(\sqrt{C}+1 \big)\pi_p \Big)^2}
\end{equation}
\end{theorem}
Intuitively, $n_u$ monotonically decreases with a decreasing $\pi_p$, and a larger $C$ in PN learning will require a larger $n_u$ to guarantee that PU learning outperforms PN learning. In practice, $\pi_p$ must be much smaller than $0.5$ because positive samples are very sparse in recommendation. As a special case, we can set $C=1$ which means the negative sampling in PN learning follows the $1:1$ balanced setting. Then, from $n_u \geq n_p/(1-2\pi_p)^2$, we easily know that when $\pi_p$ is small, e.g., less than $0.1$, PU learning is expected to outperform the corresponding PN learning with $n_u = 2n_p$. When $\pi_p$ increases, e.g., greater than $0.4$, PU learning is difficult to beat PN learning unless $n_u \geq 25n_p$. Namely, when $\pi_p$ is small (which is mostly the case for recommendation problems), PU learning is a better and computationally efficient option.

\subsection{Optimality of Convergence}
Up to now, it is still unclear whether the final convergence of PURE would enjoy the desirable property of our initial motivation of having a good generator to produce high-quality fake sample embeddings. In this section, we provide theoretical proof to show that the objective of PURE is equivalent to minimizing the KL-divergence between the true user-item relevant distribution $p_p(u,i)$ and generated distribution $p_g(u,i)$ of the generator plus unlabeled distribution $p_u(u,i)$.

First, following the analysis in~\cite{Goodfellow2014}, we show that the optimal distribution of discriminator $D$ would be a balance between $p_p(u,i)$, $p_n(u,i)$, and $p_g(u,i)$.

\begin{proposition} \label{theorem:D_optimal}
[\textbf{Optimality of the discriminator}] For a fixed generator $G$, the optimal discriminator $D$ is:
\begin{equation*}
    D^*(u,i) = \frac{\pi_p p_p(u,i)}{p_u(u,i) + p_g(u,i)}
\end{equation*}
\end{proposition}

Next, with the optimal discriminator being fixed, we can substitute $D^*(u,i)$ into the final objective of PURE in Eq.~(\ref{eq:final_objective}). Then, we can have the optimal generator as follows.

\begin{proposition}\label{theorem:G_optimal}
[\textbf{Optimality of the generator}] With the discriminator $D$ fixed, the optimization of the generator is equivalent to minimizing: $ -2\mathrm{H}\left(\frac{\pi_p}{2}\right) + \pi_p \cdot \mathrm{KL}\Big(p_{p}(u,i) \Big|\Big| \frac{p_u(u,i) +  p_{\textit{g}}(u,i)}{2}\Big) + (2-\pi_p) \cdot \mathrm{KL}\Big(\frac{(1-\pi_p)p_n(u,i)+p_{\textit{g}}(u,i)}{2-\pi_p} \Big|\Big| \frac{p_u(u,i) + p_{\textit{g}}(u,i)}{2} \Big)$ where $\mathrm{H}\left(\frac{\pi_p}{2}\right)$ is the entropy for a Bernoulli with success probability of $\frac{\pi_p}{2}$.
\end{proposition}

\begin{theorem} \label{theorem:overall_optimal}
[\textbf{Global optimum}] The global minimum could be achieved if and only if $p_p(u,i) = \frac{p_{u}(u,i)+ p_{\textit{g}}(u,i)}{2}$. At that point, the objective value of the framework $V(G,D)$ converges to $-2\mathrm{H}\left(\frac{\pi_p}{2}\right)$, and the value of $D(u,i)$ reaches $\frac{\pi_p}{2}$.
\end{theorem}

The proofs of the above four theoretical results can be found in the Appendix. In Theorem~\ref{theorem:overall_optimal}, we know the proposed framework will achieve equilibrium if and only of $p_p(u,i) = \frac{p_{u}(u,i)+ p_{\textit{g}}(u,i)}{2}$. Intuitively, upon convergence, linearly combining the optimal generator's user-item distribution with the original unlabeled user-item distribution of the given data, will be highly similar to the true relevant user-item distribution. This justifies our motivation for training a generator to produce highly relevant embeddings that confuse the discriminator as much as possible. 

\setlength{\textfloatsep}{5pt}
\begin{algorithm}[!t]
\caption{PURE}
\label{alg:PURE}
\begin{algorithmic}[1]
\State \textbf{Input:} Generators $G_u, G_i$, discriminator $D$, user-item interaction matrix $\mathcal{R}$, user set $\mathcal{U}$, item set $\mathcal{I}$, positive class prior $\pi_p$.
\State  \textbf{Initialization:} Assign $G_u, G_i$ with random weights, assign $D$ with random weights or pre-trained weights, $n_p = |\Omega|, n_u = \mathrm{ceil}\Big(\frac{n_p}{(1-2\pi_p)^2}\Big)$
\State \textbf{Repeat:}
\State \quad \textbf{for} discriminator-steps \textbf{do:}
\State \quad \quad Sample first $n_p$ tuples $(u,i) \in \Omega$ with label $1$ 
\State \quad \quad Sample another $n_p$ tuples $(u,i) \in \Omega$ with label $0$.
\State \quad \quad Sample $n_u$ tuples $(u,i) \in \Omega^-$ with label $0$.
\State \quad \quad Generate $n_u$ tuples $(u,i')$ and $(u',i)$ with label $0$ using Eq.~(\ref{generator}).
\State \quad \quad Update the discriminator model $D$ by ascending its gradients in the objectives of Eq.~(\ref{D_loss1}), and Eq.~(\ref{D_loss2}).
\State \quad \textbf{end for}
\State \quad \textbf{for} generator-steps \textbf{do:}
\State \quad \quad Generate $n_u$ random noise $z_u, z_i$ using Eq.~(\ref{eq:noise_input}).
\State \quad \quad Sample $n_u$ tuples $(u,i) \in \Omega^-$ with label $1$.
\State \quad \quad Replace $(u,i)$ with $(u,i')$ and $(u',i)$ using generator's output $G_u(z_u)$ and $G_i(z_i)$ by Eq.~(\ref{generator})
\State \quad \quad Update the generator model $G_u, G_i$ by descending their corresponding gradients in the objective of Eq.~(\ref{D_loss2}).
\State \quad \textbf{end for}
\State \textbf{Output:} The trained $G_u, G_i$, and $D$
\end{algorithmic}
\end{algorithm}

\subsection{Algorithm and Complexity}
Based on the overall learning objective, we summarize the learning steps of PURE in Algorithm~\ref{alg:PURE}. Before training, the generator and the discriminator are initialized either randomly or with pre-trained weights. Then, during the training stage, we update these two models respectively in an iterative manner. Specifically, we first fix $G$ and update the discriminator using the observed positive tuples, the sampled unlabeled tuples\footnote{We set the negative sampling ratio $C=1$ of the corresponding PN learning in the algorithm.}, and the generated user/item embeddings. Next, we fix $D$ and update the generator. The aforementioned iterative steps will continue until the model converges or the max number of iterations is reached. 

Regarding the complexity analysis of the model training, we assume both user and item have equal latent embedding dimensionality $d$. Then, the space complexity is $O\big( (M + N + 1) \cdot d \big)$ for the discriminator and is $O \big( kd\big)$ for the generator, where $k$ is the number of hidden units in generator's MLP. The computational complexity mainly involves the matrix multiplication operations. Then, the computational complexity per epoch is $O\big( (2n_p + n_u) \cdot (M + N) \cdot d^2 \big)$ for the discriminator, and $O\big( n_u k d^2 \big)$ for the generator.
\section{Experimental Results} \label{sec:experiment}
In this section, we evaluate the performance of the proposed PURE model by answering the following research questions:
\begin{description}[leftmargin=*]
    \item \textbf{RQ1:} Can the proposed PURE model outperform the state-of-the-art recommendation methods?
    \item \textbf{RQ2:} What is the parameter sensitivity for PURE in terms of positive prior $\pi_p$ and the generator's random noise input magnitude $\delta$? Does pre-train affect the ranking performance?
    \item \textbf{RQ3:} How does the running time of PURE compare with other baselines?
\end{description}

\vspace{-2mm}
\begin{table}[h!]
\setlength\tabcolsep{3pt} 
\begin{tabular}{l|c|c|c|c}
\toprule
\textbf{Dataset} & \multicolumn{1}{l|}{\textbf{\# Users}} & \multicolumn{1}{l|}{\textbf{\# Items}} & \multicolumn{1}{l|}{\textbf{\# Interactions}} & \multicolumn{1}{l}{\textbf{Sparsity}} \\ \midrule
Movielens-100k   & 943       & 1,679        & 100,000          & 6.32\%                                 \\ 
Movielens-1m     & 6,040       & 3,706       & 1,000,209       & 4.46\%                                 \\ 
Yelp        &  25,677         &   25,815       & 731,671          & 0.11\%                                 \\ \bottomrule
\end{tabular}
\caption{Statistics of the data sets}
\vspace{-10mm}
\label{dataset}
\end{table}

\subsection{Experimental Settings}
\subsubsection{Data sets.}
We conduct the experiments on three publicly accessible data sets: Movielens-100k\footnote{https://grouplens.org/datasets/movielens/100k/}, Movielens-1m\footnote{https://grouplens.org/datasets/movielens/1m/}, and Yelp\footnote{https://github.com/hexiangnan/sigir16-eals/tree/master/data}. 
For Yelp data set, due to the sparsity of the ratings among the data, we adopt the pre-processing step from ~\cite{NCF_www} by keeping the users with more than 10 item interactions. Following the experimental setting of ~\cite{IRGAN}, only the 4-star and 5-star ratings in these three data sets are treated as positive feedback, and the rest are unknown feedback. In this way, the data is transformed into the user-item interaction matrix $\mathcal{R}$ with each entry being either 0 or 1. The details of these three data sets are summarized in Table~\ref{dataset}.

\subsubsection{Baselines.}
We considered five categories of recommendation methods for comparisons:
\begin{itemize}[leftmargin=*]
    \item \textit{Traditional collaborative filtering}: \textbf{ItemPop} is a non-personalized method that recommend the most popular items to each user. \textbf{SlopeOne}~\cite{SlopeOne} infers the user-item interaction score as the sum of the user's average rating plus its average rating difference with its neighbors (who share the common items); \textbf{Co-clustering}~\cite{Co-Clustering} infers the user-item interaction score as the sum of user-item co-clustering average rating, user's average rating, and item's average rating. 
    
    \item \textit{Traditional matrix factorization}: \textbf{SVD}~\cite{SVD} infers the user-item interaction score as the sum of user bias, item bias, and the inner product of user \& item latent factors; \textbf{NMF}~\cite{NMF} is similar to SVD, but the user and item factors are computed under a non-negative constraints; \textbf{PMF}~\cite{PMF} infers the interaction score using user and item probabilistic latent factors with Frobenius regularization.
    
    \item \textit{Neural collaborative filtering}: \textbf{BPR}~\cite{BPR} learns the user and item embeddings using user-specific pairwise preferences between a pair of items; \textbf{LambdaFM}~\cite{LambdaFM} learns the embeddings using pairwise ranking loss along with lambda surrogate; \textbf{GMF}~\cite{NCF_www} learns the embeddings using pointwise label information along with relation mapping embedding. 
    
    \item \textit{GAN based recommenders}: \textbf{GraphGAN}~\cite{Wang2018} builds its discriminator to predict the connectivity between user-item vertex pair, and its generator to learn the discrete connectivity distribution; \textbf{IRGAN}~\cite{IRGAN} builds its discriminator as matrix factorization for relevance calculation, and its generator to extract discrete relevant items from the candidate pool using policy gradient~\cite{SeqGAN}; \textbf{CFGAN}~\cite{cfgan} uses the generator to generate the continuous real-valued purchase vector for each user, and the discriminator to differentiate the real purchase vectors and the generated ones.
    
    \item \textit{PU-learning based recommenders}: \textbf{PU-GMF}~\cite{nnPU} modifies the PN learning objective of GMF with its PU learning version, and feed the model with positive data and sampled unlabeled data; \textbf{PURE} is our proposed model.
    
\end{itemize}

\begin{table*}[!t]
\setlength\tabcolsep{1pt} 
\centering
\footnotesize 
\begin{tabular}{l|C{13mm}|C{13mm}|C{13mm}|C{14mm}|C{14mm}|C{14mm}|C{14mm}|C{14mm}}
\toprule
\textbf{Movielens-100k}   & \textbf{P@3} & \textbf{P@5} & \textbf{P@10} & \textbf{NDCG@3} & \textbf{NDCG@5} & \textbf{NDCG@10} & \textbf{MAP} & \textbf{MRR} \\ \midrule
\textbf{ItemPop}            &0.2624     &0.2338     &0.2049      &0.2793       &0.2568        &0.2402         &0.1515     &0.4557     \\
\textbf{SlopeOne}\cite{SlopeOne}            &0.2624    &0.2338     &0.2050      &0.2793        &0.2568        &0.2403         &0.1516     & 0.4657    \\
\textbf{Co-clustering}\cite{Co-Clustering}  &0.3158     &0.2904     &0.2471      &0.3255        &0.3088        & 0.2875        &0.1812     &0.5020     \\ \midrule
\textbf{SVD}\cite{SVD}                      &0.4028     &0.3816     &0.3289      &0.4169        &0.4037        &0.3818         &0.2522     &0.6014     \\
\textbf{NMF}\cite{NMF}                      &0.3852     &0.3509     &0.3160      &0.4025        &0.3788        &0.3655         &0.2395     &0.5925     \\
\textbf{PMF}\cite{PMF}                      &0.2624     &0.2439     &0.2068      &0.2772        &0.2620        &0.2375         &0.1462     &0.4454     \\ \midrule
\textbf{BPR}\cite{BPR}                  &0.4013     &0.3776     &0.3263      &0.4137        &0.3993       &0.3782         &0.2556     &0.6004     \\
\textbf{LambdaFM}\cite{LambdaFM}            &0.3779     &0.3496     &0.3029      &0.3966        &0.3773        &0.3575         &0.2354     &0.5897     \\ 
\textbf{GMF}\cite{NCF_www}                  &0.4042     &0.3640     &0.3156      &0.4143        &0.3897        &0.3699         &0.2524     &0.5871     \\ \midrule
\textbf{GraphGAN}\cite{Wang2018}       &0.3341 &0.2991 &0.2441 &0.3471 &0.3246 &0.2948 &0.1823 &0.5259 \\
\textbf{IRGAN}\cite{IRGAN}                  &0.4072     &0.3750     &0.314      &0.4222        &0.4009        & 0.3723        &0.2418     &0.6082     \\ 
\textbf{CFGAN} \cite{cfgan}    &0.3977 &0.3827 &0.3272 &0.4144 &0.3993 &0.3782 &0.2556 &0.6004  \\ \midrule
\textbf{PU-GMF}\cite{NCF_www}+\cite{nnPU} &0.4042 &0.3697 &0.3186 &0.4236 &0.3996 &0.3760 &0.2534 &0.6208 \\
\textbf{PURE} (ours)                              &\textbf{0.4187}     &\textbf{0.3901}     &\textbf{0.3307}      &\textbf{0.4307}        &\textbf{0.4112}        &\textbf{0.3890}         &\textbf{0.2625}     &\textbf{0.6237}     \\ \bottomrule
\end{tabular}
\caption{Evaluation results of Movielens-100k data set}
\label{results:ml-100k}
\vspace{-5mm}
\end{table*}

\begin{table*}[!t]
\setlength\tabcolsep{1pt} 
\centering
\footnotesize
\begin{tabular}{l|C{13mm}|C{13mm}|C{13mm}|C{14mm}|C{14mm}|C{14mm}|C{14mm}|C{14mm}}
\toprule
\textbf{Movielens-1m}   & \textbf{P@3} & \textbf{P@5} & \textbf{P@10} & \textbf{NDCG@3} & \textbf{NDCG@5} & \textbf{NDCG@10} & \textbf{MAP} & \textbf{MRR} \\ \midrule
\textbf{ItemPop}            &0.2805     &0.2400     &0.1845      &0.2961       &0.2725        &0.2883         &0.2371     &0.5038     \\
\textbf{SlopeOne}\cite{SlopeOne}            &0.3954     &0.3736     &0.3124      &0.3887        &0.3775        & 0.3453        &0.2958     &0.4981     \\
\textbf{Co-clustering}\cite{Co-Clustering}  &0.4826     &0.4612     &0.4195      &0.4533        &0.4475        &0.4283         &0.3500     &0.5105     \\ \midrule
\textbf{SVD}\cite{SVD}                      &0.4187     &0.3563     &0.2621      &0.4483        &0.4107        &0.4224         &0.3546     &0.6680     \\
\textbf{NMF}\cite{NMF}                      &0.5262     &0.4916     &0.4118      &0.5238        &0.5034        &0.4603         &0.4002     &0.6279     \\
\textbf{PMF}\cite{PMF}                      &0.4108     &0.3975     &0.3633      &0.3819        &0.3817        &0.3678         &0.3182     &0.4406     \\\midrule
\textbf{BPR}\cite{BPR}                  &0.6604     &0.7379     &0.8272      &0.6339        &0.6930        &0.7710         &0.3793     &0.6714     \\
\textbf{LambdaFM}\cite{LambdaFM}            &0.6365     &0.7116     &0.8072      & 0.6070       &0.6669        &0.7493         &\textbf{0.9516}     &0.6488     \\ 
\textbf{GMF}\cite{NCF_www}                  & 0.6546    &0.7284     &0.8156      &0.6254        &0.6798        &0.7583         &0.8354     &0.6594     \\ \midrule
\textbf{GraphGAN}\cite{Wang2018} &0.4731 &0.5538 &0.5209 &0.4433 &0.5072 &0.5019 &0.4198 &0.4998 \\
\textbf{IRGAN}\cite{IRGAN}                  &  0.3043   &  0.2713   &   0.2187   &    0.3225    &   0.3052     &   0.3347      &  0.2848   &   0.5441  \\
\textbf{CFGAN} \cite{cfgan}    &0.6209 &0.6978 &0.7983 &0.5902 &0.6517 &0.7379 &0.8114 &0.6337  \\ \midrule
\textbf{PU-GMF}\cite{NCF_www}+\cite{nnPU} &0.6639 &0.7394 &0.8268 &0.6398 &0.6963 &0.7724 &0.8639 &0.6762 \\
\textbf{PURE} (ours)      &\textbf{0.6824}     &\textbf{0.7523}     &\textbf{0.8351}      &\textbf{0.6532}        &\textbf{0.7094}        &\textbf{0.7829}         &0.8703     &\textbf{0.6895}      \\ \bottomrule
\end{tabular}
\caption{Evaluation results of Movielens-1m data set}
\label{results:ml-1m}
\vspace{-5mm}
\end{table*}

\begin{table*}[!t]
\setlength\tabcolsep{1pt} 
\centering
\footnotesize
\begin{tabular}{l|C{13mm}|C{13mm}|C{13mm}|C{14mm}|C{14mm}|C{14mm}|C{14mm}|C{14mm}}
\toprule
\textbf{Yelp}   & \textbf{P@3} & \textbf{P@5} & \textbf{P@10} & \textbf{NDCG@3} & \textbf{NDCG@5} & \textbf{NDCG@10} & \textbf{MAP} & \textbf{MRR} \\ \midrule
\textbf{ItemPop}            &0.1335     &0.1124     &0.0842      &0.1489       &0.1592        &0.1976         &0.1596     &0.2858     \\
\textbf{SlopeOne}\cite{SlopeOne}            &0.2053     &0.1917     &0.1567       &0.2008        &0.1983         &0.2122     &0.1986  &0.2869     \\
\textbf{Co-clustering}\cite{Co-Clustering}  &0.2216     &0.1929     &0.1475      &0.2397        &0.2499        &0.2937         &0.2423     &0.3913     \\ \midrule
\textbf{SVD}\cite{SVD}                      &0.2635     &0.2157     &0.1527      &0.2960        &0.3083        &0.3693         &0.2981     &0.4880     \\
\textbf{NMF}\cite{NMF}                      &0.3788     &0.3474     &0.2767      &0.3754        &0.3652        &0.3820         &0.3560     &0.4781     \\
\textbf{PMF}\cite{PMF}                      &0.2772     &0.2750     &0.2478      &0.2564        &0.2631        &0.2671         &0.2573     &0.3132     \\\midrule
\textbf{BPR}\cite{BPR}                  &  0.4634   & 0.5423    & 0.6561     & 0.4345       & 0.4968       &0.5918     & 0.5797     & 0.4910    \\
\textbf{LambdaFM}\cite{LambdaFM}            &0.3920     &0.4653     &0.5757      &0.3659       &0.4236        &0.5149         & 0.8173     &0.4242     \\ 
\textbf{GMF}\cite{NCF_www}                  &0.4416     &0.5230     &0.6426      & 0.4122       &0.4764        &0.5758         &0.6556     &0.4715     \\ \midrule
\textbf{GraphGAN}\cite{Wang2018} & -- &-- &-- &-- &-- &-- &-- &-- \\
\textbf{IRGAN}\cite{IRGAN}                  &  0.2643   &   0.2207  & 0.1587     & 0.2966       & 0.3136       &    0.3791     & 0.3098     & 0.4915    \\
\textbf{CFGAN} \cite{cfgan}    &0.2309 &0.2824 &0.3699 &0.2140 &0.2541 &0.3247 &0.3260 &0.2676  \\ \midrule
\textbf{PU-GMF}\cite{NCF_www}+\cite{nnPU} &0.4866 &0.5666 &0.6800 &0.4560 &0.5196 &0.6149 &0.7857 &0.5102 \\
\textbf{PURE} (ours)                              &\textbf{0.5038}     &\textbf{0.5830}     &\textbf{0.6935}      &\textbf{0.4736}        &\textbf{0.5365}        &\textbf{0.6297}         &\textbf{0.9206}     &\textbf{0.5264}    \\ \bottomrule
\end{tabular}
\caption{Evaluation results of Yelp data set}
\label{results:yelp}
\vspace{-6mm}
\end{table*}

\subsubsection{Evaluation Protocol.} To evaluate the performance of all methods, we adopt the official $80\%|20\%$ random split on Movielens-100k data set, and perform exact evaluations using the whole item sets since its size is relatively small. For Movielens-1m and Yelp data sets, we perform sampled evaluation~\cite{Koren08,NCF_www,sampled_evaluation_kdd20} to speed up the computation. In the evaluation stage, only a smaller set of random items is used as the candidates pool for ranking predictions. Due to the fact that Movielens-1m and Yelp have been pre-processed to only keep the users with at least $20$ or $10$ relevant items, we adopt the random leave-ten-out (for Movielens-1m) and leave-five-out (for Yelp) strategy to split them into the train set and the test set. Unlike the leave-one-out sampled metric strategy being used in~\cite{Koren08,NCF_www} which has candidates pool of size 100 and it may introduce bias into evaluation results. We follow the suggestion of~\cite{sampled_evaluation_kdd20} and make the candidates pool with a larger size of 500 items. This is a good trade-off pool size for the sampled evaluation where both computation cost and true performance consistency are well balanced. The eventual performance of the predicted ranked list is evaluated by Precision (P@$k$), Normalized Discounted Cumulative Gain (NDCG@$k$), where $k=\{3,5,10\}$, Mean Average Precision (MAP), and Mean Reciprocal Rank (MRR).

\subsubsection{Reproducible settings.}
\sloppy To guarantee a fair comparison between all baselines, we fix the embedding size $d$ as $5,8,16$ for all models on three data sets, respectively. Meanwhile, the input allowed for all models would be the rating matrix $\mathcal{R}$ only, no side information or additional features are supplied. All models are validated on the performance of P@5. The learning rate is searched from $\{1 \times 10^{-4}, 1 \times 10^{-3}, 1 \times 10^{-2}\}$, the positive class prior $\pi_p$ is searched from $\{1\times 10^{-6},1\times 10^{-5},1\times 10^{-4},1\times 10^{-3},1\times 10^{-2},1\times 10^{-1}\}$, the generator's input noise magnitude $\delta$ is searched from $\{0.001, 0.005, 0.01, 0.05, 0.1, 0.2, 0.3\}$. Training is accelerated with pre-train, where we initialize PURE's generator with random weights and initialize the discriminator with PU-GMF's embedding weights.

\subsection{Performance Comparison (RQ1)}
From the evaluation results shown in Table~\ref{results:ml-100k}, Table~\ref{results:ml-1m}, and Table~\ref{results:yelp}, we observe that PURE achieves the best performance on most metrics over these three data sets. The neural collaborative filtering methods and GAN-based methods are usually very competitive overall. Regarding the results of Movielens-100k in Table~\ref{results:ml-100k}, we see that there exists at least one baseline per category that performs relatively well since it is a small-scale and well-preprocessed data set. Among all competitive baselines with good results, PURE can outperform over SVD, BPR, IRGAN, and CFGAN with $1\%-2\%$ on average. 

For the Movielens-1m data set, as we can see in Table~\ref{results:ml-1m}, BPR, LambdaFM, PU-GMF, and CFGAN also perform relatively well in terms of P@$k$ and NDCG@$k$ comparing with other baselines. One interesting observation is that LambdaFM has very high values on the MAP metric. That is because pairwise learning is position-independent and pairwise-ordering at the bottom of the ranking list would impact the learning loss as much as the top pairs. Meanwhile, LambdaFM is particularly designed for optimizing the overall ranking performance. We follow the setting from the papers of BPR and IRGAN to tune the learning rate, number of epochs, etc. We found that BPR is very sensitive to the sample sequence in the training batches, and IRGAN's performance is largely impacted by its hyperparameters and pretraining. We have performed a comprehensive model tuning in a reasonable time period for all baselines and reported their best performance for a fair comparison.

For the results of the Yelp data set, the best frameworks are PU-GMF and PURE, followed by neural collaborative filtering methods such as BPR, LambdaFM, and GMF. The GraphGAN method fails to finish training on Yelp since it needs to compute the graph softmax and generate a huge amount of neighbor vertices for each existing vertex. The traditional collaborative filtering and MF-based methods do not have a satisfactory performance on Yelp. We also observe that GAN-based methods could easily fail to converge even with careful hyperparameter tuning. The reasons for their poor performance are two-fold: First, these methods are taking the unobserved data as negative samples without the negative sampling procedure, resulting in an unbalanced training data problem, especially for Yelp data set, which is much sparser than the other two data sets.  
Second, they didn't use continuous space sampling, and generating with discrete sampling will end up with poor model expressiveness especially when dealing with large-scale sparse data set. As a comparison, PU learning will help sample from both the observed and unobserved entries and the generator will further learn the data distribution and generate continuous user-item embeddings to increase the model expressiveness. Similarly, the models with pairwise loss (BPR and LambdaFM) also perform relatively well on MAP due to their position-independent properties in the modeling. 

\begin{figure}[!t]
\centering
\vspace{-3mm}
\includegraphics[width=6.1cm]{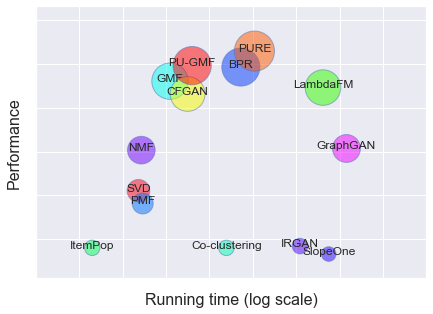}
\vspace{-4mm}
\caption{Running time of baselines on Movielens-1m}
\label{run_time}
\end{figure}

\vspace{-2mm}
\subsection{Parameter Study (RQ2)}
Regarding the hyper-parameters sensitivity in PURE, we show the performance of P@5, NDCG@5, MAP, and MRR with respect to the positive class prior $\pi_p$ and the magnitude of generator's input noise $\delta$. We perform this parameter study on the Movielens-100k data set because of its small size so that we can well-tune all the competitive baselines with a reasonable amount of effort. 

\begin{figure*}[!t]
\centering
\includegraphics[width=16cm]{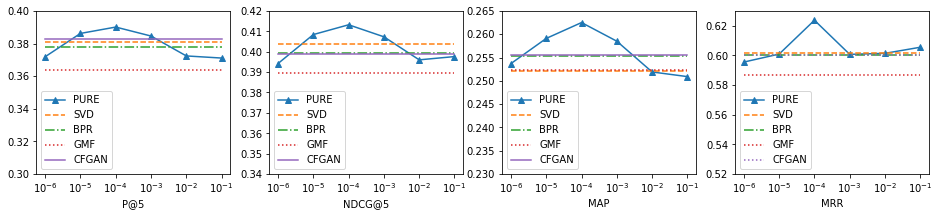}
\vspace{-4mm}
\caption{Positive class prior $\pi_p$}
\vspace{-3mm}
\label{pos_ratio}
\end{figure*}

First, we can see that PURE achieves the best performance with carefully selected hyper-parameters. We observe in Figure~\ref{pos_ratio} that PURE outperforms (on average $1.5\%$) all the competitors when $\pi_p$ is set to $0.0001$. Meanwhile, we see that PURE has a performance guarantee if $\pi_p$ falls into the range of $[0.00001, 0.001]$ which means the underlying true density of the positive samples is very sparse. Namely, each user would only show interest in a very small number of items on average, which is reasonable in real-world applications. 

\begin{figure*}[!t]
\centering
\includegraphics[width=16cm]{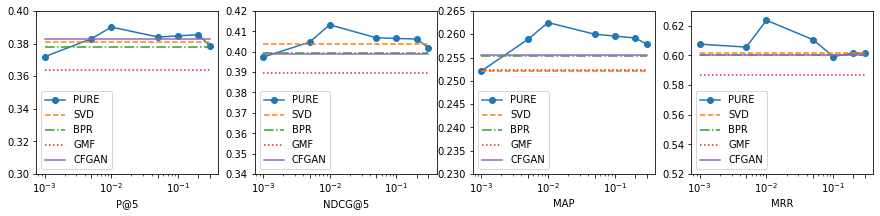}
\vspace{-4mm}
\caption{Generator's input noise magnitude $\delta$ }
\vspace{-3mm}
\label{emb_std}
\end{figure*}

Second, Figure~\ref{emb_std} shows that PURE is not very sensitive to the conditional noise magnitude. Starting from $\delta = 0.005$ to $\delta=0.1$, we observe that PURE can almost outperform every baseline in all metrics. It is because the generator could produce high-quality fake embeddings to help improve the discriminating ability of the discriminator. We conjecture that fine-tuning the structure of MLP layers could further improve the expressiveness of the generator, which in turn improves the overall performance. The exploration of the optimal model structure is left for future work.

Third, to demonstrate the utility of pre-training, we compared the performance of two different versions for PURE - with and without pre-training. For PURE without pre-training, we initialize the embedding layers of the discriminator and the MLP layers of the generator with random weights. For PURE with pre-train, we first train a PU-GMF model, and then assign its embedding weights to PURE's discriminator, but the generator is still random initialized. As shown in Table.~\ref{pretrain_compare}, the relative improvements of utilizing pre-training are roughly $1\%$, $4\%$, and $5\%$ for Movielens-100k, Movielens-1m, and Yelp, respectively. We empirically observe that PURE with pre-training converges faster with less training epochs. Both above observations justify the usefulness and efficiency of our proposed pre-training method for initializing PURE.



\begin{table}[!t]
\setlength\tabcolsep{4pt} 
\centering
\small
\begin{tabular}{l|c|c|c|c}
\toprule
\multirow{2}{*}{}       & \multicolumn{2}{c|}{\textbf{With Pretrain}} & \multicolumn{2}{c}{\textbf{Without Pretrain}} \\ \cmidrule{2-5} 
                        & \textbf{P@5}        & \textbf{NDCG@5}       & \textbf{P@5}         & \textbf{NDCG@5}         \\ \midrule
\textbf{Movielens-100k} & \textbf{0.3901}              & \textbf{0.4112}               & 0.3833               & 0.4094                  \\ 
\textbf{Movielens-1m}   & \textbf{0.7523}              & \textbf{0.7094}                & 0.7101               & 0.6568                  \\ 
\textbf{Yelp}           & \textbf{0.5830}              & \textbf{0.5365}                & 0.5340               & 0.4864                  \\ \bottomrule
\end{tabular}
\caption{Performance of PURE with/without pre-training.}
\vspace{-4mm}
\label{pretrain_compare}
\end{table}

\vspace{-2mm}
\subsection{Running Time (RQ3)}
In Figure~\ref{run_time}, we compare the running time between PURE and other baselines on Movielens-1m. The circle size represents the average performance (NDCG@5 in this case) of the corresponding method and the X-axis records their running time in log scale. As we can see, the traditional matrix factorization methods run very fast but with limited performance. PURE performs the best but a little slower than PU-GMF due to the extra training time of the generator. The pairwise loss based models, such as BPR and LambdaFM, are comparable in terms of performance. CFGAN performs well on this data set since it also performs continuous sampling. IRGAN suffers the issue of high computational complexity due to multiple reasons, e.g., dynamic negative sampling, softmax operations, high generator, and discriminator epochs. GraphGAN and SlopeOne need to loop over all users and items multiple times and therefore, have the highest computational complexity.

\section{Related Work} \label{sec:relatedwork}
\subsection{Recommender Systems}
Algorithms and frameworks regarding recommendation systems have been widely studied in recent years due to their business success to attract traffic or improve profit in different domains~\cite{Jannach2019}. Collaborative filtering based methods play an important role in recommender systems and gain major attention~\cite{SVD} for recent decades. Within collaborative filtering, latent factor or embedding based algorithms such as matrix factorization~\cite{SVD}, factorization machines~\cite{FM} and their variants~\cite{NMF,PMF, LambdaFM} have been successfully applied in recommender systems. With the development of deep neural networks, deep learning based recommender systems become a hot research topic since they introduce non-linearity and increase model expressiveness~\cite{Zhang2019}. Traditional matrix factorization based algorithms have been transformed into their deep model versions, such as neural collaborative filtering~\cite{NCF_www} and deep factorization machine \cite{deepfm}. Wide \& Deep model combines the traditional linear model with extensive features and the deep model in order to trade off between memorization and generalization \cite{Cheng2016, NFM}. 
In order to approach the true distribution of the user and items, generative adversarial networks have been adopted for information retrieval \cite{IRGAN} and network mining \cite{HeGAN,Wang2018}. IRGAN~\cite{IRGAN} formulates a minimax game where the generator learns the discrete relevance distribution of users and items for synthesizing the indistinguishably fake user-item tuples while the discriminator identifies whether one user-item tuple is real or not. Following this idea, GraphGAN~\cite{Wang2018} learns the underlying connection distribution over vertices in an adversarial framework for graph representation learning. HeGAN~\cite{HeGAN} further proposed the relation-aware generator and discriminator to encode the heterogeneous information network with multiple types of vertices and edges.

All these methods take the unlabeled user and item interactions as negative samples, which is a non-valid assumption for real-world applications. In this paper, we addressed this problem by utilizing the GAN-based retrieval model and train its discriminator under the PU learning framework. 

\subsection{Positive Unlabeled Learning}
PU learning is a variant of the classical PN learning, where the training data only consists of positive and unlabeled samples. This learning setting fits with the applications that do not require fully supervised data, e.g., one-class learning~\cite{Bernhard1999} and semi-supervised learning~\cite{SSL}. The pioneering work \cite{LetouzeyDG00, ComiteDGL98} of PU learning was initialized two decades ago. 
The state-of-the-art PU learning approaches are mainly focused on unbiased PU risk estimators. Starting from~\cite{ElkanN08}, which treats the unlabeled data as a weighted mixture of positive and negative data and has an unbiased estimator if positive and negative conditional densities are disjoint, multiple variants~\cite{nnPU, PUtheory, PlessisNS14} have been proposed. It has been proven in~\cite{PlessisNS14} that an unbiased PU estimator can be learned if the loss is symmetric. Later on, the analysis in~\cite{PUtheory} shows that the unbiased estimator could be convex for loss functions that meet the linear-odd condition. However, the aforementioned approaches are not applicable to very flexible models, where the overall risk of the estimator will become negative. ~\cite{nnPU} has shown that by imposing a non-negative operator on the estimated empirical risk term of the unlabeled data, the non-negative risk estimator will reduce the overfitting phenomenon, which opens the door for adopting deep neural networks into PU learning frameworks. 

PU learning has not been extensively explored on recommender systems, even inherently the recommendation problem fits the PU learning scenario very well. Most related work includes PU learning for matrix completion~\cite{Hsieh2015}, and positive-unlabeled demand-aware recommendation \cite{Yi2017}, which is performing a tensor completion with a low-rank assumption. Our work is different from these methods in that, PURE does not impose any assumption on the distribution of users and items, and employs the GAN framework to learn the real distribution of the user-item interaction in a continuous embedding space.

\section{Conclusion} \label{sec:concludsion}
In this paper, we proposed a novel recommendation framework named PURE based on the generative adversarial network. The discriminator of PURE is trained using PU learning with an unbiased risk estimator, while the generator learns the underlying continuous distribution of users and items in order to generate high-quality fake embeddings of them. We theoretically analyzed the performance of PURE from multiple aspects, and empirically performed extensive experiments to demonstrate its effectiveness and efficiency for personalized ranking problems in comparison with a rich set of strong baselines.

\bibliographystyle{ACM-Reference-Format}
\bibliography{acmart.bib}

\appendix
\onecolumn

\section{Appendix} 
\noindent \textbf{In all the following proofs, we denote each user-item tuple $(u,i)$ in recommendation as one data sample $x$ for simplicity}.

\subsection{Proof of Theorem~\ref{theorem:sampling_bound}}
Theorem~\ref{theorem:sampling_bound} states that the Estimation Error Bound of PU learning is tighter than that of PN learning if and only if:
\begin{equation*} \label{eq:PUsampling_appendix}
    n_u \geq \frac{\sqrt{C} \, n_p}{\Big(1-\big(\sqrt{C}+1 \big)\pi_p \Big)^2}
\end{equation*}

\begin{proof}
The differences of PN learning and PU learning in terms of the EER bounds in Lemma~\ref{lemma:eeb} reflect the differences w.r.t. their risk minimizers. We define:
\begin{equation}
    \alpha_{pu,pn} := \frac{\pi_p/\sqrt{n_p} + 1/\sqrt{n_u}}{(1-\pi_p)/\sqrt{n_n}}
\end{equation}
For simplicity, let's denote $\rho_{pu} := n_p/n_u$ and we know $\rho_{pn} := n_p/n_n = 1/C$. Then, by setting $\alpha_{pu,pn} \leq 1$, we have:
\begin{equation}
    \begin{split}
        \pi_p + \sqrt{\rho_{pu}} \leq \frac{1}{\sqrt{C}}(1-\pi_p) \Leftrightarrow &\; \rho_{pu} \leq \frac{1}{\sqrt{C}}\Big(1-(\sqrt{C}+1)\pi_p\Big)^2 
    \end{split}
\end{equation}
It is rather straightforward to get the conclusion in Eq.~(\ref{eq:PUsampling_appendix}) by solving the above inequality.
\end{proof}

\begin{lemma} \label{lemma:eeb}
[\textbf{Estimation Error Bound (EEB)}] Let $\mathcal{F}$ be the function class, and $\hat{f}_{pn}$ and $\hat{f}_{pu}$ be the empirical risk minimizer of $\,\hat{R}_{pn}(D)$ and $\hat{R}_{pu}(D)$ for discriminator $D$ that belongs to PN learning and PU learning, respectively. Then, the EEB of $\hat{f}_{pu}$ is tighter than $\hat{f}_{pn}$ with probability at least $1-\delta$ when:
\begin{equation} \label{eq:PUcondition}
    \frac{\pi_p}{\sqrt{n_p}} + \frac{1}{\sqrt{n_u}} < \frac{\pi_n}{\sqrt{n_n}}
\end{equation}
if the loss $L$ is symmetric and Lipschitz continuous, and the Rademacher complexity of $\mathcal{F}$ decays in $O(1/\sqrt{n})$ for data of size $n$ drawn from $p_{data}(x), p_p(x)$, and $p_n(x)$.
\end{lemma}
Proof can be referred to~\cite{PUtheory} for details. Based on the above theorem, we know that PU learning is highly likely to outperform PN learning when Eq.~(\ref{eq:PUcondition}) and certain mild conditions~\cite{PUtheory} are satisfied.

\subsection{Proof of Proposition~\ref{theorem:D_optimal}}
Proposition~\ref{theorem:D_optimal} states that when the generator $G$ is fixed, the optimal discriminator $D$ is:
\begin{equation*}
    D^*(x) = \frac{\pi_p p_p(x)}{p_u(x) + p_{g}(x)}
\end{equation*}
\begin{proof}
We know that the underlying true data distribution is: $p_{data}(x) = \pi_p p_p (x) + (1-\pi_p) p_n(x)$. Furthermore, we also denote the generator's output distribution as $p_g(x)$. Then, the objective of the discriminator $D$ is as follows for fixed $G$:
\begin{equation} \label{eq:D_obj_in_proof}
    \begin{aligned}
        \max_{D} V(D) &= \pi_p \int_x p_p(x) \log(D(x))dx - \pi_p \int_x p_p(x)\log(1-D(x)) dx  + \int_x p_u(x) \log(1-D(x)) dx +  \int_z p_z(z) \log(1-D(G(z))) dz \\
        &= \pi_p \int_x p_p(x) \log(D(x))dx - \pi_p \int_x p_p(x)\log(1-D(x)) dx + \int_x p_u(x) \log(1-D(x)) dx +  \int_x p_g(x) \log(1-D(x)) dx \\
        &= \pi_p \int_x p_p(x) \log(D(x))dx + \int_x \Big( - \pi_p\cdot p_p(x) + p_u(x) +  p_g(x) \Big) \log(1-D(x)) dx \\
        &= \int_x \pi_p \cdot p_p(x) \log(D(x))dx + \int_x \Big( (1 - \pi_p)\cdot p_n(x) +  p_g(x) \Big) \log(1-D(x)) dx
    \end{aligned}
\end{equation}
Here, we assume that the unlabeled distribution can also be decomposed as $p_{u}(x) \approx \pi_p p_p (x) + (1-\pi_p) p_p(x)$ approximately since the sampled positive tuples are extremely sparse in the overall user-item population. Next, we know that for the problem of $\max_{y} a \log(y) + b\log(1-y)$, it achieves the optimal value~\cite{Goodfellow2014} when $y^*=\frac{a}{a+b}$. Let $a=\pi_p \cdot p_p(x)$ and $b=(1 - \pi_p)\cdot p_n(x) + \cdot p_g(x)$,
\begin{equation*}
    D^* = \frac{\pi_p \cdot p_p(x)}{\pi_p \cdot p_p(x) + (1 - \pi_p)\cdot p_n(x) +  p_g(x)} = \frac{\pi_p \cdot p_p(x)}{p_u(x) +  p_g(x)}
\end{equation*}
\end{proof}

\subsection{Proof of Proposition~\ref{theorem:G_optimal}}
Theorem~\ref{theorem:G_optimal} states that when the discriminator $D$ fixed, the optimization of the generator $G$ is equivalent to minimize: $ -2H\left(\frac{\pi_p}{2}\right) + \pi_p \cdot \mathrm{KL}\Big(p_{p}(x) \Big|\Big| \frac{p_u(x) +  p_{\textit{g}}(x)}{2}\Big) + (2-\pi_p) \cdot \mathrm{KL}\Big(\frac{(1-\pi_p)p_n(x)+p_{\textit{g}}(x)}{2-\pi_p} \Big|\Big| \frac{p_u(x) + p_{\textit{g}}(x)}{2} \Big)$.
\begin{proof}
For fixed optimal discriminator, we substitute $D^*$ into the objective of Eq.~(\ref{eq:D_obj_in_proof}) and have the following objective:
\begin{equation*}
    \begin{aligned}
        & \min V(G) = \pi_p \int_x p_p(x) \log(D(x))dx - \pi_p \int_x p_p(x)\log(1-D(x)) dx + \int_x p_u(x) \log(1-D(x)) dx + \int_z p_z(z) \log(1-D(G(z))) dz \\
        &= \int_x \pi_p  p_p(x) \log(D(x))dx + \int_x \Big( (1 - \pi_p) p_n(x) + p_g(x) \Big) \log(1-D(x)) dx \\
        &= \int_x \pi_p  p_p(x) \log\left(\frac{\pi_p  p_p(x)}{p_u(x) + p_g(x)}\right)dx + \int_x \Big( (1 - \pi_p) p_n(x) +  p_g(x) \Big) \log\left(1-\frac{\pi_p  p_p(x)}{p_u(x) +  p_g(x)}\right) dx \\
        &= \int_x \pi_p  p_p(x) \left[\log \frac{\pi_p}{2} + \log\left(\frac{p_p(x)}{\frac{p_u(x) + p_{\textit{g}}(x)}{2} }\right)\right]dx  + (2-\pi_p)\int_x \frac{(1 - \pi_p) p_n(x) +  p_g(x)}{2-\pi_p} \left[ \log\left(\frac{\frac{(1 - \pi_p) p_n(x) + p_{\textit{g}}(x)}{2 - \pi_p} }{\frac{p_u(x) + p_{\textit{g}}(x)}{2} }\right) + \log \frac{2-\pi_p}{2}\right]dx \\
        &= \pi_p \log \frac{\pi_p}{2} + \pi_p \cdot \mathrm{KL}\left( p_p(x) \Big|\Big| \frac{p_u(x) + p_g(x)}{2} \right) + (2-\pi_p)\log \frac{2-\pi_p}{2} + (2-\pi_p) \cdot \mathrm{KL}\left(\frac{(1-\pi_p)p_n(x)+p_g(x)}{2-\pi_p} \Big|\Big| \frac{p_u(x) + p_g(x)}{2} \right) \\
        &= -2\mathrm{H} \left(\frac{\pi_p}{2}\right) + \pi_p \cdot \mathrm{KL}\left(p_{p}(x) \Big|\Big| \frac{p_u(x) +  p_g(x)}{2}\right) + (2-\pi_p) \cdot \mathrm{KL}\left(\frac{(1-\pi_p)p_n(x)+p_g(x)}{2-\pi_p} \Big|\Big| \frac{p_u(x) + p_g(x)}{2} \right)
    \end{aligned}
\end{equation*}

\end{proof}

\subsection{Proof of Theorem~\ref{theorem:overall_optimal}}
Theorem~\ref{theorem:overall_optimal} states that the global minimum could be achieved if and only if $p_p(x) = \frac{p_{u}(x)+ p_{\textit{g}}(x)}{2}$. At that point, the objective value of the framework $V(G,D)$ converges to $-2\mathrm{H}\left(\frac{\pi_p}{2}\right)$, and the value of $D(x)$ reaches $\frac{\pi_p}{2}$.

\begin{proof}
From Proposition~\ref{theorem:G_optimal}, we can directly get the minimum of the optimal generator as $-2 \mathrm{H}(\frac{\pi_p}{2})$ if and only if these three distributions are identical: $p_{p}(x) = \frac{p_u(x) +  p_{\textit{g}}(x)}{2}$ and $ \frac{(1-\pi_p)p_n(x)+p_{\textit{g}}(x)}{2-\pi_p} = \frac{p_u(x) +  p_{\textit{g}}(x)}{2}$. By solving the second equality, we have:
\begin{equation}
\begin{aligned}
    & \frac{p_u(x) + p_g(x)}{2} = \frac{(1-\pi_p)p_n(x)+p_g(x)}{2-\pi_p} \\ \Longleftrightarrow \quad & (2-\pi_p) p_u(x) + (2-\pi_p) p_g(x) = (2-2\pi_p)p_n(x) + 2p_g(x) \\
    \Longleftrightarrow \quad & (2-\pi_p)p_u(x) - \pi_p p_g(x) = (2-2\pi_p) p_n(x)  \\
    \Longleftrightarrow \quad & (2-\pi_p)p_u(x) -\pi_p p_g(x) = 2 p_u(x) - 2\pi_p p_p(x) \quad \big(\mathrm{by \, substituting} \; (1-\pi_p) p_n(x) = p_u(x)- \pi_p p_p(x) \big)\\
    \Longleftrightarrow \quad & p_p(x) = \frac{p_u(x) + p_g(x)}{2}
\end{aligned}
\end{equation}
which is exactly the same as the first equality. Then, if we substitute either of them into the optimal $D^*$, we will always have $D^*(x) = \frac{\pi_p}{2}$.

\end{proof}









\subsection{Parameter Setting}
To recover the experimental results, below are the required reproducible settings: For all three data sets, we trained the generator from scratch with ``lecun\_uniform'' random initialization on the MLP layers. For discriminator, we initialize its user and item embedding weights with a pre-trained PU-GMF weights.
The model losses for both discriminator and generator are binary cross entropy loss and they are optimizer using Adam optimizer. The local epochs for the discriminator and the generator are $1$ and $10$, respectively. The MLP layer in generator has {\tt ReLU} activation which has been verified to perform better than other activation functions, such as {\tt LeakyReLU}, {\tt Sigmoid}, {\tt Linear}, etc. Other detailed hyperparameter settings are summarized in Table~\ref{reproducibility}. 


\begin{table}[h!]
\setlength\tabcolsep{3pt} 
\begin{tabular}{l|c|c|c|c|c|c}
\toprule
&\textbf{latent dim. $d$}  &\textbf{batch size} & \textbf{learning rate} & \textbf{\# epoch} & \textbf{pos. prior $\pi_p$} & \textbf{noise mag. $\delta$}  \\ \midrule
\textbf{Movielens-100k}&5  &128 &0.001  &100  &0.0001   &0.01  \\ 
\textbf{Movielens-1m}  &8  &128 &0.001  &100  &0.00001  &0.01   \\   
\textbf{Yelp}          &16 &512 &0.001  &200  &0.000001 &0.01   \\ \bottomrule  
\end{tabular}
\caption{Reproducible parameter setting}
\label{reproducibility}
\vspace{-8mm}
\end{table}


\end{document}